\documentclass[twocolumn,5p]{autart} 

\usepackage{graphicx}          

\usepackage{tikz}
\usetikzlibrary{shapes}
\usepackage{enumitem}
\usepackage{placeins}
\usepackage{amsmath,amssymb,mathtools,stmaryrd}
\usepackage{algorithmic}
\usepackage{graphicx}
\usepackage{url}
\usepackage{multirow}

\newtheorem{remark}{Remark}
\newtheorem{lemma}{Lemma}
\newtheorem{theorem}{Theorem}
\newtheorem{proposition}{Proposition}
\newtheorem{corollary}{Corollary}

\usepackage{bm}
\usepackage{bbm}

\DeclareMathOperator{\diag}{diag}

\newcommand{\bse}{\begin{subequations}}
\newcommand{\ese}{\end{subequations}}

\newcommand{\be}{\begin{equation}}
\newcommand{\ee}{\end{equation}}

\newcommand{\bbm}{\begin{bmatrix}}
\newcommand{\ebm}{\end{bmatrix}}

\newcommand{\re}[1]{\mathbb{R}^{#1}}%

\usepackage{flushend}

\begin{document}

\begin{frontmatter}

\title{Zonotope-Based Elastic Tube Model Predictive Control\thanksref{footnoteinfo}} 

\thanks[footnoteinfo]{Corresponding author: Sabin Diaconescu.}

\author[UPB]{Sabin Diaconescu}\ead{sabin.diaconescu@stud.acs.upb.ro},
\author[UPB]{Florin Stoican}\ead{florin.stoican@upb.ro},
\author[UPB]{Bogdan D. Ciubotaru}\ead{bogdan.ciubotaru@upb.ro},
\author[SUPELEC]{Sorin Olaru}\ead{sorin.olaru@centralesupelec.fr}

\address[UPB]{Department of Automatic Control and Systems Engineering, \\ Nat. Univ. of Science and Technology Politehnica Bucharest, 060042, Bucharest, Romania.}
\address[SUPELEC]{Laboratoire des Signaux et Syst\`emes, CentraleSup\'elec, Paris-Saclay University, Gif-sur-Yvette 91192, France.}


\begin{abstract} 
Tube-based Model Predictive Control (MPC) is a widely adopted robust control framework for constrained linear systems under additive disturbance. The paper is focused on reducing the numerical complexity associated with the tube parameterization, described as a sequence of elastically-scaled zonotopic sets. A new class of scaled-zonotope inclusion conditions is proposed, alleviating the need for a priori specification of certain set-containment constraints and achieving significant reductions in complexity. A comprehensive complexity analysis is provided for both the polyhedral and the zonotopic setting, illustrating the trade-off between an enlarged domain of attraction and the required computational effort. The proposed approach is validated through extensive numerical experiments.
\end{abstract}
\end{frontmatter}

\section{Introduction} \label{sec:intro}
\vspace{-1em}
Model Predictive Control (MPC) is extensively utilized in both industrial practice and academic research \cite{Diehl2017,Mayne2014}. It offers a versatile framework that enables simultaneous performance optimization and constraint satisfaction, while systematically balancing computational demands with cost-function (sub)optimality \cite{Diehl2017,Borrelli2017}. Mathematical models of real systems are inherently approximate and seldom capture the underlying dynamics with complete accuracy. This limitation, well recognized in the literature \cite{Morari1999}, has motivated extensive research in robust and stochastic MPC \cite{Cannon2015}. \emph{Tube-based MPC} tackles model uncertainties and exogenous disturbance by employing a sequence of sets to ensure that all future trajectories of the system are contained therein (see \cite{Blanchini2008set}).\vspace{-1em}

The numerical tractability of the tube-based MPC optimization problem is typically ensured via set parameterizations. The rigid tube MPC approach uses a fixed-shape translation to form the tube \cite{Mayne2005,Rakovic2023}, while the homothetic tube MPC method introduces scalable sets \cite{Rakovic2012}. The parameterized tube MPC represents tube cross-sections as convex hulls of finitely many points, incurring quadratic growth in decision variables w.r.t. to the horizon length \cite{RakovicParameterized2012}. Alternatively, the elastic tube framework employs a fixed number of half-spaces in the set description, via an elasticity parameter \cite{Fleming2015,RakovicElastic2016} that controls each half-space's offset. The main benefits of the homothetic and elastic approaches are the enlarged domain of attraction and improved performance. The sets constituting the tube are commonly represented as polyhedra; however, their complex descriptions significantly increase the size of the MPC problem and complicate its implementation. Ellipsoidal representations \cite{Parsi2022} provide an alternative, though they tend to be conservative and introduce quadratic constraints. Configuration-constrained tube MPC methods \cite{Villanueva2024}, compute the control input at each tube vertex. 
\vspace{-1em}

Zonotopes have been increasingly applied in recent studies across diverse domains, including set estimation~\cite{rego2020guaranteed}, fault detection~\cite{raimondo2016closed}, and tube MPC formulations~\cite{Andrade2024,raghuraman2021tube,Raghuraman2022Hierarhical,KoelnReduced2023}. Scaled zonotopes have also been employed to define tube profiles~\cite{ioan2019navigation,yang2021scalable}, differing from earlier approaches~\cite{Rakovic2012,RakovicElastic2016} by enforcing constant scaling factors throughout the prediction horizon. The uncertainty sets in~\cite{raghuraman2021tube,Raghuraman2022Hierarhical,KoelnReduced2023} are adjustable, with generator scaling used to minimize the size of robust positively invariant sets~\cite{Raghuraman2022Sets}. In contrast, \cite{How2006} iteratively computes feedforward reachable sets to bound uncertainties, controlling complexity through a closed-loop nilpotent state matrix. Although \cite{raghuraman2021tube} employs a tube-based construction, it maintains fixed scaling factors and considers adjustable uncertainty sets.

This work extends the authors' preliminary work on elastic tube MPC \cite{DiaconescuEtAlCDC24}. We introduce a novel linear encoding for zonotope containment that generalizes standard conditions, thereby circumventing the need to preselect auxiliary variables. These results extend naturally to scaled zonotopes, eliminating the bilinearities typically associated with scaling factors and avoiding suboptimal parameter selection. Furthermore, this formulation enables a computationally efficient precomputation-based alternative. The Robust Positively Invariant (RPI) set approximation is also enhanced by treating the zonotope centers as independent decision variables. The resulting elastic-tube MPC problem is reformulated to allow for an explicit quantification of computational trade-offs. In contrast to polyhedral scaling methods that involve bilinear terms, the proposed inclusion test remains linear in both auxiliary variables and scaling factors, facilitating its integration into large-scale optimization frameworks. Finally, conditions ensuring recursive feasibility and asymptotic stability are established.

Additionally, necessary and sufficient conditions for zonotopic robust control tubes were recently established in \cite{Houska2025} using vertex control. While these conditions offer a theoretically exact characterization of feasible tubes, they necessitate verifying constraints at every vertex of the zonotope. Combinatorial vertex growth makes high-order zonotope verification computationally prohibitive. In contrast, the proposed method leverages scalable linear encodings that circumvent vertex enumeration entirely, prioritizing numerical tractability. The number of vertices for a zonotope embedded in $\mathbb{R}^n$ with $D$ generators is upper-bounded by $2 \sum_{i=0}^{n-1} \binom{D-1}{i}$ \cite[Thm. 9.4]{fukuda2020polyhedral}, rendering vertex-based methods impractical for higher-dimensional systems.

The paper is organized as follows: Section~\ref{sec:prereq} introduces zonotope and tube MPC fundamentals. Section~\ref{sec:rmpc_zonotope} presents the scaled zonotopic tube parameterization and recursive feasibility analysis. Complexity and practical applications are detailed in Sections~\ref{sec:complexity} and \ref{sec:apps}, followed by conclusions in Section~\ref{sec:conclusions}.


\noindent\emph{Notation}. The set of integers in interval $[x_1, x_2]\subset \mathbb{R}$ is denoted by $\mathcal{I}_{[x_1,x_2]}$. If $x_2=\infty$ the shortcut $\mathcal{I}_{\geq x_1} := \mathcal{I}_{[x_1,\infty]}$ is used. The terms $\mathbf{0}_n$ and $\mathbf{1}_n$ denote the length-$n$ zero and one vectors, respectively. For two sets $P,Q\subset \mathbb{R}^n$, their Minkowski sum and Pontryagin difference are $P \oplus Q = \{\, p+q : p\in P, q\in Q \,\}$ and $P \ominus Q = \{\, p : \{p\}\oplus Q \subseteq P \,\}$. A polyhedron is represented in half-space form as $X = \{\, x \in \mathbb{R}^n : Fx \leq \theta \,\}$, where $F \in \mathbb{R}^{q\times n}$, $\theta \in \mathbb{R}^q$, where $q$ is the number of constraints. The symbol $\otimes$ denotes the Kronecker product. The term $x_{m:n}$ denotes the sequence $\{x_m,\ldots,x_n\}$ for $m,n \in \mathcal{I}_{\geq 0}$. $X_{(i,j)}$ denotes the matrix entry on row $i$ and column $j$. $X_{(i,:)}$ denotes the $i$-th row and $X_{(:, j)}$, the $j$-th row.  $X \geq Y$ denotes element-wise inequality $X_{(i,j)} \geq Y_{(i,j)}$.

\section{Preliminaries and problem formulation} \label{sec:prereq}
A zonotope \cite{fukuda2020polyhedral} is a centrally symmetric polytope that can be expressed as the Minkowski sum of line segments. In its \textit{generator representation}, it is defined as
\be
\label{eq:z}
Z=\langle c,G\rangle=\{x\in \mathbb R^n:\: x=c+G\xi, \: \| \xi \|_{\infty} \leq 1 \},
\ee 
where $c \in \re{n}$ denotes its \textit{center}   and $G\in \mathbb R^{n\times D}$ its \textit{generator matrix}. The zonotopes are \cite{fukuda2020polyhedral}: closed under affine transformations, closed under Minkowski sum, symmetric up to their center.
A natural generalization is the concept of a scaled zonotope \cite{yang2021scalable}, defined by expressing $G \leftarrow G\Delta$, where $\Delta=\diag(\delta)$, with $\delta\in \mathbb R^D_{\geq 0}$.


Consider the linear time-invariant (LTI) system
\begin{equation}
\label{eq:discrete_LTI_0}
    x_{k+1}= A x_k+B u_k+\omega_k,
\end{equation}
where $x_k\,, x_{k+1}\in\mathcal X=\langle c_X, G_X\rangle\subset \mathbb{R}^n$ denote the current and successor state, $u_k\in\mathcal U=\langle c_U, G_U\rangle\subset \mathbb{R}^m$ the control input, and $\omega_k\in \mathcal W=\langle c_W, G_W\rangle\subset \mathbb{R}^{n}$ the exogenous disturbance; the matrices $A\in\mathbb{R}^{n\times n}$, $B\in\mathbb{R}^{n\times m}$ are the known state and input matrices, while $\mathcal X$, $\mathcal U$ and $\mathcal W$ denote the sets bounding their associated signals. Since any zonotope may be equivalently represented in its polyhedral half-space form, \cite[Sec. 4.8]{fukuda2020polyhedral}, let us denote $\mathcal X=\{x\in \mathbb R^n:\: F_Xx\leq \theta_X\}$, $\mathcal U=\{u\in \mathbb R^m:\: F_Uu\leq \theta_U\}$ and $\mathcal W=\{\omega\in \mathbb R^{n}:\: F_W\omega\leq \theta_W\}$.

Let $\bar x_k$ denote the nominal state trajectory. The tuple $(\bar x_k, c_k, \delta_k)$ is chosen such that the set membership holds
\begin{equation}
    \label{eq:tuple}
    x_k \in \bar x_k \oplus \langle c_k, G\Delta_k\rangle.
\end{equation}
\begin{defn}
    For $k\in \mathcal I_{N-1}$, the sequence $(\bar x_{k}, c_{k}, \delta_{k})$ checking \eqref{eq:tuple} is called admissible w.r.t. \eqref{eq:discrete_LTI_0} if there exists $u_k \in \mathcal U$ such that, for all $\omega_k \in \mathcal W$, the following holds:
    \begin{enumerate}[label=\roman*),nosep]
        \item one-step feedforward reachability
        \begin{multline}
        \label{eq:one-step}
            x_k\in \bar x_k \oplus \langle c_k, G\Delta_k\rangle \implies \\
            x_{k+1}\in \bar x_{k+1} \oplus \langle c_{k+1}, G\Delta_{k+1}\rangle, \: k\in \mathcal I_{N-1};
        \end{multline}
        \item state admissibility
        \begin{equation}
        \label{eq:state-adm}
            x_{k} \in \bar x_{k} \oplus \langle c_{k}, G\Delta_{k}\rangle \subseteq\langle c_X, G_X\rangle, \: k\in \mathcal I_N.
        \end{equation}
    \end{enumerate}
\end{defn}
Additionally, input admissibility reduces to verifying
\begin{equation}
    \label{eq:input-adm}
    u_k\in \langle c_U, G_U\rangle,\, k\in \mathcal I_{N-1}.
\end{equation}
\begin{proposition}
    \label{prop:admissibility}
   Under the control action
    \begin{equation}
    \label{eq:u}
        u_k =\bar u_k + K(x_k-\bar x_k),
    \end{equation}
    for an admissible sequence $(\bar x_k, c_k, \delta_k)$, $k\in \mathcal I_{N-1}$, 
    the one-step reachability implication \eqref{eq:one-step} holds if
        \begin{multline}
        \label{eq:one-step-condition}
            \left\langle A\bar x_k+B\bar u_k+A_Kc_k , A_KG\Delta_k\right\rangle\oplus\langle c_W, G_W\rangle\\ \subseteq \langle \bar x_{k+1}+ c_{k+1}, G\Delta_{k+1}\rangle \,,
        \end{multline}
        and admissibility w.r.t. \eqref{eq:state-adm}--\eqref{eq:input-adm} is ensured by
        \begin{align}
            \label{eq:state-condition}\langle \bar x_k +c_k, G\Delta_k\rangle &\subseteq \langle c_X, G_X\rangle,\\
            \label{eq:input-condition}\langle \bar u_k +Kc_k, KG\Delta_k\rangle &\subseteq \langle c_U, G_U\rangle,
        \end{align}
        provided that a suitable nominal control sequence $\{\bar u_k\}_{k \in \mathcal I_{N-1}}$ can be found.
\end{proposition}

\begin{pf}
Taking the tuple $(\bar x_k,c_k,\delta_k)$ from \eqref{eq:tuple} and using \eqref{eq:u}, the constraints \eqref{eq:state-condition}--\eqref{eq:input-condition}, involving the nominal state and input together with their disturbance sets, ensure the admissibility requirements
\eqref{eq:state-adm}--\eqref{eq:input-adm}.
Substituting $u_k$ into \eqref{eq:discrete_LTI_0}, the successor state must satisfy
\begin{multline*} 
x_{k+1}\in \left(A\bar x_k+B\bar u_k\right)\oplus A_K \langle c_k, G\Delta_k\rangle \oplus\langle c_W, G_W\rangle\,, 
\end{multline*}
with the shorthand notation $A_K = A+BK$. Defining $(\bar x_{k+1},c_{k+1},\delta_{k+1})$ as in \eqref{eq:tuple}, condition \eqref{eq:one-step-condition} ensures \eqref{eq:one-step}. Repeating for each $k\in \mathcal I_{N-1}$ concludes the proof. 
\hfill\hfill \qed
\end{pf}

For the initial state $x_0\in \mathcal X_0\subseteq \mathcal X$ and the predicted input sequence $\mathbf u_{N}=\{u_0,u_1,\ldots, u_{N-1}\}$, whose components are a combination of feedforward ($\bar u_k$) and feedback action ($K(x_k-\bar x_k)$), the constrained-optimization problem\footnote{The nominal dynamics are not enforced as strict equality constraints.} 
\begin{subequations}
\label{eq:MPC_problem_nominal}
\begin{align}
    \label{eq:mpc_nominal_a}\bar{\mathbf u}^\star_N=\arg\min_{\bar{\mathbf u}_N, } V_T(\bar x_N, \delta_N)&+\sum\limits_{k=0}^{N-1} \ell(\bar x_k, \bar u_k, \delta_k),\\
    \label{eq:mpc_nominal_b0}\text{s.t.}\,\: 
    x_0 \in \mathcal X_0 &\subseteq \bar x_0\oplus \langle c_0, G\Delta_0\rangle,\\ 
    \label{eq:mpc_nominal_b1}\eqref{eq:one-step-condition}-\eqref{eq:input-condition},\: k &\in \mathcal{I}_{[0,N-1]} \,,\\
    \label{eq:mpc_nominal_d} \{\bar x_N\}\oplus \langle c_N, G\Delta_N \rangle& \subseteq \mathcal T \subseteq \mathcal{X},
\end{align}
\end{subequations}
ensures the reachability and admissibility conditions \eqref{eq:one-step}--\eqref{eq:input-adm}. Over a prediction horizon of length $N$, the nominal state evolves according to \eqref{eq:one-step-condition}. The stage states and inputs are constrained by bounds \eqref{eq:state-condition} and \eqref{eq:input-condition}, while the terminal state is restricted to a terminal set \eqref{eq:mpc_nominal_d}, satisfying the standard recursive feasibility properties \cite{Diehl2017}. 

The cost function, defined in \eqref{eq:mpc_nominal_a}, detailed later in Section~\ref{subsec:feasibility}, penalizes a stage cost and a terminal cost, both depending on the nominal state/input and scaling factors. The control input $u_k$, applied to the system \eqref{eq:discrete_LTI_0}, is constructed by taking the first element of the optimal control sequence $\bar{\mathbf u}^\star_N\in\mathbb{R}^{Nm}$, to which the feedback term $K(x_k-\bar x_k)$ is appended.

This tube-MPC formulation, adapted from \cite{Mayne2005,Houska2019} and first introduced in \cite{DiaconescuEtAlCDC24}, compares favorably to standard polyhedral implementations \cite{RakovicElastic2016}. The core novelty of \eqref{eq:MPC_problem_nominal} lies in the use of scaled zonotopic sets, which tighten the original constraints and allow efficient set containment formulations in \eqref{eq:one-step-condition}--\eqref{eq:input-condition}. 
The goal of the current work is to refine the approach from \cite{DiaconescuEtAlCDC24} by alleviating the need to preselect auxiliary variables, imposed to circumvent a nonlinear optimization problem in the polyhedral/zonotopic case. The next section lays the foundation for this development.

\section{Robust MPC with scaled zonotopic sets} \label{sec:rmpc_zonotope}
This section revisits the polyhedral framework of \cite{RakovicElastic2016} and employs it as a ground to introduce the elastic zonotopic approach, for which subsequent sections develop the constructive and computational details.

\subsection{Scaled zonotope inclusion conditions}
\label{subsec:scaled_zon_inclusion}
The next lemma restates a classical result, presented first in \cite[Cor. 4]{sadraddini2019linear}, and generalizes it to dimensions $n>2$.

\begin{lemma}
\label{lem:zon_inclusion}
Let there be zonotopic sets $Z_i=\langle c_i, G_i\rangle\subset \mathbb R^d$, with $c_i\in \mathbb R^d, G_i\in \mathbb R^{d\times D_i}$, for $i\in \mathcal I_n$. A sufficient condition for the inclusion $Z_1\oplus \ldots \oplus Z_{n-1}\subseteq Z_n$ is the existence of matrices $\Gamma_i \in \mathbb R^{D_n\times D_i}$, for $i\in \mathcal I_{n-1}$, and of vector $\gamma \in \mathbb R^{D_n}$, verifying
\begin{subequations}
\label{eq:zon_inclusion_test}
\begin{align}
    \label{eq:zon_inclusion_test_a}
    G_i=G_n\Gamma_i,\quad \sum\limits_{i=1}^{n-1}c_i&=c_n+G_n\gamma,\\ \label{eq:zon_inclusion_test_b}\sum\limits_{i=1}^{n-1}|\Gamma_i|\mathbf 1_{D_i}+ |\gamma|&\leq \mathbf 1_{D_n}.
\end{align}
\end{subequations}
\end{lemma}
\begin{pf}
    Let there be some $x_i\in Z_i$. By definition, this implies the existence of $\lambda_i\in \mathbb R^{D_i}$ such that $x_i = c_i + G_i\lambda_i$ and $\|\lambda_i\|_\infty\leq 1$. The following chain of implications holds
    \begin{align*}
        x&=\sum\limits_{i=1}^{n-1} \left(c_i+G_i\lambda_i\right)\overset{\eqref{eq:zon_inclusion_test_a}}{=}c_n + G_n\gamma+\sum\limits_{i=1}^{n-1} G_n\Gamma_i\lambda_i\\
        &= c_n + G_n\left(\gamma + \sum\limits_{i=1}^{n-1}\Gamma_i\lambda_i\right).
    \end{align*}
Clearly, $\|\gamma + \sum\limits_{i=1}^{n-1}\Gamma_i\lambda_i\|_\infty\leq 1$ implies that $x\in Z_n$. Using both the norm's triangle inequality and the initial assumption of $\|\lambda_i\|_\infty\leq 1$ leads to \eqref{eq:zon_inclusion_test_b}. \hfill\hfill\qed 
\end{pf}

\begin{proposition}
\label{prop:zon_inclusion_test_phi}
   Conditions \eqref{eq:zon_inclusion_test_a}-\eqref{eq:zon_inclusion_test_b} may be written equivalently as
\begin{subequations}  
\label{eq:zon_inclusion_test2}
    \begin{align}    
    \label{eq:zon_inclusion_test2_a}
    G_n\Phi^\top V^\top &= \bbm\left(-c_n + \sum\limits_{i=1}^{n-1}c_i\right) & G_1 & \ldots \, G_{n-1}\ebm,\\
    \label{eq:zon_inclusion_test2_b}\Phi& \geq \mathbf 0_{2\bar D\times D_n},\quad \Phi^\top\mathbf 1_{2\bar D}\leq \mathbf 1_{D_n},
    \end{align}
\end{subequations}
with $V=\bbm I_{\bar D} & -I_{\bar D}\ebm$, $\Phi\in \mathbb R^{2\bar D \times D_n}$, and $\bar D=1+\sum\limits_{i=1}^{n-1}D_i$.
\end{proposition}
\begin{pf}
The $j$-th component of the element-wise inequality \eqref{eq:zon_inclusion_test_b} can be rewritten as 
\begin{equation}
\label{eq:j-component}
    |\gamma_{(j)}|+\sum_{i=1}^{n-1}\sum_{\ell=1}^{D_i}|\Gamma_{i, (j,\ell)}|\leq 1.
\end{equation}
By denoting $\xi_{(j,:)} = \bbm \gamma_{(j)} & \Gamma_{1, (j,:)} & \ldots & \Gamma_{n-1, (j,:)}\ebm$, \eqref{eq:j-component} reduces to $\|\xi_{(j,:)}\|_1\leq 1$, which is equivalent with the inclusion $\xi_{(j,:)}^\top\in \mathcal{CR}_{\bar D}$. The unit cross-polytope $\mathcal{CR}_{\bar D}\subset\mathbb R^{\bar D}$ is the polar set of the hypercube \cite{fukuda2020polyhedral} and it is notorious for its large number of facets ($2^{\bar D}$). On the other hand, it has only $2\bar D$ vertices, so it becomes appealing to introduce auxiliary variables and describe the inclusion $\xi_{(j,:)}^\top\in \mathcal{CR}_{\bar D}$ as a convex sum of vertices
\begin{equation}
\label{eq:Gamma_CR2}
    \xi_{(j,:)} = \Phi_{(:,j)}^\top V^\top ,\: \Phi_{(:,j)}\geq \mathbf 0_{2\bar D},\: \Phi_{(:,j)}^\top\mathbf 1_{2\bar D} \leq 1,
\end{equation}
where $\Phi_{(:,j)}$ is the $j$-th column of the matrix of coefficients $\Phi \in \mathbb R^{2\bar D \times D_n}$. Enumerating and stacking in \eqref{eq:Gamma_CR2} for $j\in \mathcal I_{D_n}$ and further noting that the equalities from \eqref{eq:zon_inclusion_test_a} may be regrouped to put in evidence $\xi$, one obtains
\begin{align*}
&\xi = \Phi^\top V^\top, \quad \Phi \ge \mathbf 0_{2\bar D\times D_n}, \quad \Phi^\top\mathbf 1_{2\bar D} \le \mathbf 1_{D_n},\\
&\bbm\left(-c_n + \sum_{i=1}^{n-1}c_i\right)
\; G_1 \; \ldots \; G_{n-1}\ebm = \\
& = G_n
\bbm \gamma \; \Gamma_1 \; \ldots \; \Gamma_{n-1}\ebm
= G_n\xi.
\end{align*}
Introducing $\xi$ from the first equality into the second one leads to \eqref{eq:zon_inclusion_test2_a}-\eqref{eq:zon_inclusion_test2_b}.\hfill\hfill\qed
\end{pf}
\vspace{-1em}

Next, consider the case of scaled zonotopes.
\begin{corollary}
\label{cor:scaled_zon_condition}
Let there be scaled zonotopic sets $Z_i(\delta_i)=\langle c_i, G_i\diag(\delta_i)\rangle\subset \mathbb R^d$, with $c_i\in \mathbb R^d, G_i\in \mathbb R^{d\times D_i}$, $\delta_i\in \mathbb R^{D_i}_+$, for $i\in \mathcal I_n$. A sufficient condition to verify the inclusion $Z_1(\delta_1) \oplus \, \ldots \, \oplus Z_{n-1}(\delta_{n-1})\subseteq Z_n(\delta_n)$ is the existence of the coefficient matrix $\Phi \in \mathbb R^{2\bar D \times D_n}$, verifying
\begin{subequations}  
\label{eq:zon_inclusion_test3}
    \begin{align}    
    \label{eq:zon_inclusion_test3_a}
    \mkern-14mu G_n\Phi^\top V^\top\mkern-8mu &=\mkern-4mu \mkern-4mu \bbm\mkern-4mu\left(\mkern-4mu -c_n \mkern-4mu +\mkern-4mu \sum\limits_{i=1}^{n-1}c_i\mkern-4mu\right)& G_1\Delta_1& \ldots G_{n-1}\Delta_{n-1}\ebm,\\
    \label{eq:zon_inclusion_test3_b}\Phi& \geq \mathbf 0_{2\bar D\times D_n},\quad \Phi^\top\mathbf 1_{2\bar D}\leq \delta_n \,.
    \end{align}
\end{subequations}
\end{corollary}
\begin{pf}
Replacing $G_i$ with $G_i\Delta_i$, $i \in  \mathcal{I}_{n}$, in \eqref{eq:zon_inclusion_test2} results in
\begin{subequations}
\begin{align}
    \label{eq:proof_cor_scaled_1}
    \mkern-8mu G_n\Delta_n\Phi^\top V^\top\mkern-4mu &\mkern-8mu=\mkern-8mu \bbm-c_n \mkern-4mu+\mkern-6mu \sum\limits_{i=1}^{n-1}\mkern-4mu c_i& G_1\Delta_1& \ldots G_{n-1}\Delta_{n-1}\ebm\mkern-4mu,\\
    \label{eq:proof_cor_scaled_2}
    \mkern-28mu \Phi& \geq \mathbf 0_{2\bar D\times D_n},\quad \Phi^\top\mathbf 1_{2\bar D}\leq \mathbf 1_{D_n}.
\end{align}
\end{subequations}
Right- and left-multiplying \eqref{eq:proof_cor_scaled_2} with the element-wise positive  $\Delta_n$ leads to 
$\Phi\Delta_n \geq \mathbf 0_{2\bar D\times D_n}\cdot \Delta_n,\quad \Delta_n\Phi^\top\mathbf 1_{2\bar D}\leq \Delta_n\cdot\mathbf 1_{D_n}$.
Recalling that $\Delta_n\mathbf 1_{D_n}=\delta_n$ and replacing $\Phi$ with $\Phi\Delta_n$ leads to \eqref{eq:zon_inclusion_test3}.
\hfill\hfill\qed
\end{pf}

\begin{remark}
    Noteworthily, \eqref{eq:zon_inclusion_test3} is linear in both the variable $\Phi$ and the scaling factors $\delta_i, i\in \mathcal I_n$. Thus, it can be easily embedded in larger optimization problems requiring zonotope inclusions checks. This is not the case of polyhedral scaling, as discussed in \cite{RakovicElastic2016}, where bilinear terms appear in the equivalent form of \eqref{eq:zon_inclusion_test3_b}. \hfill \hfill \qed
\end{remark}

The issue, especially when embedding \eqref{eq:zon_inclusion_test3} into a prediction horizon mechanism, is that, at each step, $2\bar D\times D_n$ new auxiliary variables are introduced. This is addressed, sub-optimally, in the next result.
\begin{corollary}
\label{cor:Phi0}
Consider the notation from Cor.~\ref{cor:scaled_zon_condition} and $\Phi_0$ which verifies conditions \eqref{eq:zon_inclusion_test2}. Then, 
\begin{equation}
\label{eq:Phi_0_cond}
    \Phi_0^\top \left(\bbm 1\\1\ebm\otimes \bbm 1\\ \delta_1\\\vdots\\ \delta_{n-1}\ebm\right)\leq \delta_n
\end{equation}
is a sufficient condition for checking inclusion $Z_1(\delta_1)\oplus \ldots \oplus Z_{n-1}(\delta_{n-1})\subseteq Z_n(\delta_n)$.
\end{corollary}
\begin{pf}
Denoting $T=\diag\left(1, \delta_1, \ldots, \delta_{n-1}\right)$, right-multiplying with it in \eqref{eq:zon_inclusion_test2_a} and noting that $V^\top T = (I_2 \otimes T)\cdot V^\top$, one obtains 
\begin{equation}
\label{eq:Phi_zero_mod}
    \Phi=(I_2 \otimes T)\Phi_0\Delta_n^{-1},
\end{equation}
which verifies \eqref{eq:proof_cor_scaled_1}. Further, $\Phi_0\geq \mathbf 0_{2\bar D\times D_n}$ implies that $\Phi\geq \mathbf 0_{2\bar D\times D_n}$, since $T,\Delta_{n}^{-1}$ are element-wise positive. What remains to be checked is that $\Phi^\top\mathbf 1_{2\bar D}\leq \mathbf 1_{D_n}$ holds. With notation \eqref{eq:Phi_zero_mod}, this becomes
\begin{equation}
    \Delta_n^{-1}\Phi_0^\top (I_2 \otimes T)\mathbf 1_{2\bar D} \leq \mathbf 1_{D_n},
\end{equation}
which, recalling that $\Delta_n\geq 0$ and that $\Delta_n\mathbf 1_{D_n}=\delta_n$, $T\mathbf 1_{\bar D}=\bbm 1& \delta_1^\top & \ldots & \delta_{n-1}^\top \ebm^\top$, directly leads to \eqref{eq:Phi_0_cond}.
\hfill\hfill\qed
\end{pf}

\begin{remark}
\label{rem:phi0}
The matrix $\Phi_0$ can be pre-computed by solving an optimization problem, analogous to \cite[Prop. 2]{RakovicElastic2016},
\begin{equation}
\label{eq:Phi_0_computation}
\mkern-4mu
    \Phi_0 = \arg\min_\Phi \|\Phi^{\top} \cdot \bm 1_{2\bar D}\|_\infty \enskip \text{s.t. \eqref{eq:zon_inclusion_test2}}\,.
\end{equation}
This eliminates, at runtime, the need to enforce \eqref{eq:zon_inclusion_test3}, replacing it with the linear constraint \eqref{eq:Phi_0_cond}. As a result, no additional auxiliary variables are introduced into the main optimization problem. The drawback, however, is that fixing the value of $\Phi_0$ may lead to an infeasible problem for certain combinations of scaling factors that would otherwise remain feasible under \eqref{eq:zon_inclusion_test3}. A similar pre-computation strategy was employed in \cite{RakovicElastic2016} for scaled polyhedral sets. \hfill \hfill \qed
\end{remark}

\subsection{Computation of the RPI set}
The inclusion conditions \eqref{eq:one-step-condition}--\eqref{eq:input-condition} rely on the seed-generator matrix $G$. Although it could be chosen randomly, it is common practice \cite{RakovicElastic2016} to derive it from the Robust Positive Invariant (RPI) set associated with the error dynamics $z_{k+1}= A_K z_k + \omega_k, \: \omega_k\in \mathcal W$, where $z_k = x_k - \bar x_k$. This is a prerequisite for rigid tubes and improves the feasibility of the problem for the homothetic and elastic cases. 

Containment conditions for zonotopic sets \cite{raghuraman2021tube,sadraddini2019linear} can be employed to compute a one-step zonotopic minimal RPI (mRPI) approximation. The following result extends \cite[Prop. 4]{DiaconescuEtAlCDC24} by leveraging Cor. \ref{cor:scaled_zon_condition} and explicitly treating the center as a decision variable.
\begin{proposition}
\label{prop:rpi_approx_new}
Let $\mathcal W = \left<c_{\omega}, G_{\omega} \right>$ and construct $\bar G = \bbm I & A_K & \ldots & A_K^{s-1}\ebm G_{\omega} \in \mathbb R^{n\times D}$, for some fixed positive integer $s$. The following linear program
\begin{subequations}
\label{eq:zonotopic_rpi}
    \begin{align}
        \label{eq:cross_1_new}\min_{c, \Phi, \delta} & \quad\delta\\
        \label{eq:cross_2_new}\bar G \Phi^{\top} V^{\top} &= \bbm -c + A_K c + c_{\omega} & A_K \bar G \Delta & G_{\omega} \ebm, \\
        \label{eq:cross_3_new} \Phi &\geq \mathbf 0_{2\bar D\times D}, \: \Phi^{\top} \mathbf 1_{2\bar D} \leq \delta\,,
    \end{align}
\end{subequations}
provides the center $c\in \mathbb R^n$ and the scaling factor $\delta\in \mathbb R^{D}$ such that the zonotopic set $\left<c,G=\bar G\Delta\right>$ is RPI, with shorthand notation $\bar{D} = D + D_{\omega} + 1$.
\end{proposition}

\begin{pf}
A candidate set $\left<c,\bar G\Delta\right>$ is RPI under dynamics $z_{k+1}= A_K z_k + \omega_k$ iff set inclusion $A_K\left<c,\bar G\Delta\right>\oplus \mathcal W\subseteq \left<c,\bar G\Delta\right>$ holds. The program \eqref{eq:zonotopic_rpi} results from Cor.~\ref{cor:scaled_zon_condition} by taking $Z_1=\left<A_K c, A_K \bar G \Delta\right>$, $Z_2=\left<c_{\omega}, G_{\omega} \right>$ and $Z_3 = \left<c, \bar G \Delta \right>$, where $\Delta=\diag(\delta)$.\hfill\hfill\qed
\end{pf}
\vspace{-3mm}
The center $c$ is folded into the disturbance set's center ($c_\omega \leftarrow c_\omega + c$) to avoid cluttering the subsequent computations.

\begin{remark}
\label{rem:mRPI_Rakovic}
     The mRPI set can be approximated arbitrarily close
     \cite[Thm. 1]{rakovic_invariant_2005} by $F(\mu,s) = (1-\mu)^{-1}\bigoplus_{i=0}^{s-1}A_K^{i}\mathcal W,$ where, for a fixed positive integer $s$, $0<\mu<1$ is the smallest scalar verifying $A_K^s\mathcal W\subseteq \mu \mathcal W$. For a given $s$, if $\mu<1$, it is guaranteed that \eqref{eq:zonotopic_rpi} is feasible with a possible solution being $c=\left(I - A_K\right)^{-1} c_{\omega}$, and $\delta = (1-\mu)^{-1}\mathbf 1_{D}$. The additional degrees of freedom due to the independent generator scaling in \eqref{eq:zonotopic_rpi} means that, for the same index $s$, $\left<c, \bar G \Delta \right> \subseteq F(\mu,s)$. While not explicitly enforced, a constraint such as $\delta \leq (1-\mu)^{-1}\mathbf{1}_{D}$ can be imposed to ensure inclusion whenever a $\mu$ exists for a given $s$. \hfill \hfill \qed
\end{remark}

\begin{remark} \label{rem:complexity_vs_CDC}
Denoting by $D_{\omega}$ the number of generators in  $G_{\omega}$, the linear program of Prop.~ \ref{prop:rpi_approx_new} involves $2sD_{\omega}(1 + D_{\omega} + s D_{\omega})$ decision variables, fewer than the $3 sD_{\omega} (1 + D_{\omega} + s D_{\omega})$ variables used in \cite[Prop. 4]{DiaconescuEtAlCDC24}. \hfill \hfill \qed
\end{remark}

\begin{remark} \label{rem:zonotope_vs_polyhedron}
    Under hyper-rectangular constraints, the mRPI is a zonoid and thus ideally suited for zonotopic approximation. Without symmetry, zonotopic effectiveness diminishes, necessitating polyhedral descriptions that typically compromise performance.
\end{remark}

\subsection{Recursive feasibility analysis}
\label{subsec:feasibility}
The recursive feasibility and stability topics in the context of tube MPC are described in more detail in \cite{Mayne2005,RakovicElastic2016,raghuraman2021tube,Mayne2000}. Hereinafter, the classical construction of \cite{ChenAllgower1998} is considered, in the context of \cite{RakovicElastic2016}, and adapted for the elastic case. With the notation from \eqref{eq:MPC_problem_nominal}, consider the stage and terminal costs  
\begin{align}
    \nonumber \ell(\bar x, \bar u, \delta)&=\bar x^\top Q_x \bar x + \bar u^\top Q_u\bar u + (\delta-\mathbf 1_D)^\top Q_\delta (\delta-\mathbf 1_D),\\
    \label{eq:cost-recursive-feasibility}V_T(\bar x, \delta)&=\bar x^\top P_x \bar x + (\delta-\mathbf 1_D)^\top P_\delta (\delta-\mathbf 1_D),
\end{align}
with $Q_x, Q_\delta\succeq 0$, $Q_u\succ 0$ given and $P_x=P^\top_x \succeq 0, P_\delta=P^\top_\delta \succ 0$ to be obtained subsequently. Summing over the prediction horizon yields the total cost 
\begin{equation}
\label{eq:cost}
    V_N(x_0)= V_T(\bar x_N, \delta_N)+\sum\limits_{k=0}^{N-1} \ell(\bar x_k, \bar u_k, \delta_k)\,.
\end{equation}

Assume a terminal control law
\begin{equation}
\label{eq:control_terminal}
    \bar u_T(\bar x)=K_T\bar x,
\end{equation} 
with an associated terminal set $\mathcal T \subseteq \mathcal{X}$, computed as the Maximal Positive Invariant (MPI) set. In the context of the total cost \eqref{eq:cost}, the recursive feasibility and asymptotic stability properties are guaranteed by verifying:
\begin{enumerate}[label=\roman*)]
    \item robust set invariance
    \begin{multline}
    \label{eq:terminal_invariance}
        \exists \, \{\bar{x}, c, \delta\}_{N+i}, \{\bar{x}, c, \delta\}_{N+i + 1}\, \\
        \text{ s.t.} \,\,
        x_{N+i} \in \mathcal T
        \implies
        x_{N+i+1}\in \mathcal T,\, 
        \forall i \in \mathcal{I}_{\geq 0};
    \end{multline}
    \item state and input admissibility conditions
    \begin{equation}
    \label{eq:terminal_admissibility}
        x_{N+i}\in \mathcal X, \, u_{N+i}\in \mathcal U, \, \forall i \in \mathcal{I}_{\geq 0};
    \end{equation}
    \item Lyapunov-like contraction
    \begin{multline}
    \label{eq:terminal_contractiveness}
        V_T(\bar x_{N+1}, \delta_{N+1})-V_T(\bar x_N, \delta_N)\\+\ell (\bar x_N, K_T\bar x_N, \delta_N) \leq 0.
    \end{multline}
    \end{enumerate}
Introducing \eqref{eq:control_terminal} in \eqref{eq:discrete_LTI_0} allows to recast \eqref{eq:terminal_admissibility}, via the constructions from \eqref{eq:one-step-condition}--\eqref{eq:input-condition}, into the equivalent relations
\begin{subequations}
\label{eq:extended-constraints}
   \begin{align}
     \label{eq:extended-constraints_a}
     F_X(\bar x_{N+i} + c_{N+i})+|F_XG|\delta_{N+i} &\leq \mathbf 1 \,,\\
     \label{eq:extended-constraints_b}
     F_U(K_T\bar x_{N+i}+Kc_{N+i})+|F_UKG|\delta_{N+i}& \leq \mathbf 1,
    \end{align}
\end{subequations}
which have to be respected for all subsequent iterations of the closed-loop extended dynamics
\begin{subequations}
\label{eq:extended-dynamics}
   \begin{align}
    \label{eq:extended-dynamics_a}\bar x_{N+i+1}&=A_{K_T}\bar x_{N+i}, \\
    \label{eq:extended-dynamics_b}\delta_{N+i+1}&=\Phi_0^\top \left(\mathbf 1_2\otimes \bbm \mathbf 1_{1+D_{\omega}}\\ \delta_{N+i}\ebm\right)=L\delta_{N+i} + d \,.
\end{align}
\end{subequations}
Following the reasoning in \cite{RakovicElastic2016}, several simplifying assumptions were adopted to reach this representation: the set inclusion \eqref{eq:one-step-condition} becomes an equality, yielding \eqref{eq:extended-dynamics_a}, while \eqref{eq:Phi_0_cond} serves to both provide $\Phi_0$ and to arrive at \eqref{eq:extended-dynamics_b}. The matrices $L$ and $d$ provide a compact representation of the right-hand side of the middle equality in \eqref{eq:extended-dynamics_b}, after rearranging columns to highlight $\delta_{N+i}$. Henceforth, the set-recurrence \cite[Thm. 3]{rakovic2022implicit}
\begin{multline}
\label{eq:mpi_rec_standard}
    \Omega_0=\bigl\{(\bar x_N^\top, \delta_N^\top)^\top:\: \text{s.t. }\eqref{eq:extended-constraints_a}-\eqref{eq:extended-constraints_b}\text{ hold}\bigr\}, \\ \Omega_{k+1}=\bbm A_{K_T}&0\\0& L\ebm^{-1}\left(\Omega_k \oplus \bbm0\\-d\ebm\right)\cap \Omega_0,
\end{multline}
is guaranteed to terminate at some finite index $\bar k$, thus enabling the choice $\mathcal T=\{(\bar x,\delta):\: F_T(\bar x,\delta)\leq \theta_T\} = \Omega_{\bar k}$, which satisfies \eqref{eq:terminal_invariance}. Lastly, substituting \eqref{eq:extended-dynamics} into \eqref{eq:terminal_contractiveness} yields the following Linear Matrix Inequalities (LMIs)
\begin{subequations}
\label{eq:terminal_cost_cond}
    \begin{align}
        \label{eq:terminal_cost_cond_a}A_{K_T}^\top P_xA_{K_T}-P_x+Q_x+K_T^\top Q_u K_T&\prec 0,\\
        \label{eq:terminal_cost_cond_b}L^\top P_\delta L - P_\delta + Q_\delta&\prec 0,
    \end{align}
\end{subequations}
which ensure that \eqref{eq:terminal_contractiveness} holds. It is often convenient to enforce equality, reformulating \eqref{eq:terminal_cost_cond} as algebraic Riccati equations, thereby enabling the analytic computation of $P_x, P_\delta$.

\subsection{The scaled zonotope elastic tube optimization}
By combining the preceding results, \eqref{eq:MPC_problem_nominal} can be reformulated into the main result of the paper -- an efficient solution for an elastic-tube MPC problem employing a scaled zonotopic tube profile.
\begin{theorem}
\label{thm:elastic-tube-Phi}
For dynamics \eqref{eq:discrete_LTI_0} under feedback law \eqref{eq:u}, $G$ obtained as in Prop.~\ref{prop:rpi_approx_new} and $\mathcal T, P_x,P_\delta$ determined from \eqref{eq:mpi_rec_standard}--\eqref{eq:terminal_cost_cond}, the constrained quadratic-programming (QP) optimization problem\footnote{Notation $\cdot_{\mathbf N}$ is a shorthand for sequence $\{\cdot_k\}_{\mathcal I_{[0, N-1]}}$.}
\begin{subequations}  
    \label{eq:zon_incl_compact}
    \begin{align}
    \label{eq:zon_incl_compact_a}\bar{\mathbf u}^\star_{\mathbf N}=\arg\mkern-12mu\min_{\mkern-40mu\bar{\mathbf u}_{\mathbf N}, \Phi_{\mathbf N}, \delta_{\mathbf N},\bm c_{\mathbf N}, \bm \lambda_{\mathbf N}}\mkern-24mu V_T(\bar x_N, \delta_N)+\sum\limits_{k=0}^{N-1} &\ell(\bar x_k, \bar u_k, \delta_k),\\
    \label{eq:zon_incl_compact_b}\text{s.t.}\: 
    \bar x_0\oplus \langle c_0, G\Delta_0\rangle &\supseteq \mathcal X_0, \\
    \label{eq:zon_incl_compact_c}c_k =G\lambda_k, \: |\lambda_k|&\leq \delta_k,\\
    \label{eq:zon_incl_compact_d}F_X(c_k+\bar x_k)+\left|F_XG\right|\delta_k&\leq \theta_X,\\
    \label{eq:zon_incl_compact_e}F_U(Kc_k+\bar u_k)+\left|F_UKG\right|\delta_k&\leq \theta_U,\\
    \label{eq:zon_incl_compact_i}F_T(c_N+\bar x_N,\,\delta_N)&\leq \theta_T,\\
    \label{eq:zon_incl_compact_g} G \Phi_{k}^\top V^\top = \bigl[-c_{k+1}-\bar x_{k+1}+ A\bar x_k& + B \bar u_k \\ 
    \nonumber+ A_K c_k + c_{\omega}\bigr|& \: A_K G \Delta_k \:\:\bigr| G_{\omega} \bigr],\\
    \label{eq:zon_incl_compact_h} \Phi_{k} \geq \mathbf 0_{2\bar D\times D},\: \Phi_{k}^\top\mathbf 1_{2\bar D}&\leq \delta_{k+1},
    \end{align}
    \end{subequations}
    for all $k \in \mathcal{I}_{[0,N-1]}$, verifies admissibility conditions \eqref{eq:one-step}--\eqref{eq:input-adm} and recursive feasibility guarantees \eqref{eq:terminal_invariance}--\eqref{eq:terminal_contractiveness}.
\end{theorem}
\vspace{-2em}
\begin{pf}
Constraints \eqref{eq:zon_incl_compact_d}--\eqref{eq:zon_incl_compact_e} ensure state and input admissibility, as defined in \eqref{eq:state-adm}--\eqref{eq:input-adm}. Constraint \eqref{eq:zon_incl_compact_i} checks \eqref{eq:mpc_nominal_d}, with $\mathcal T$ obtained from set recurrence \eqref{eq:mpi_rec_standard}. Constraints \eqref{eq:zon_incl_compact_c} ensure that $c_k \in \langle 0, G \Delta_k\rangle$. Considering Cor.~\ref{cor:scaled_zon_condition} and making the required notations (for the case $n=3$) allows to recast the one-step condition \eqref{eq:one-step} into \eqref{eq:zon_incl_compact_g}--\eqref{eq:zon_incl_compact_h}. Thus, the sequences of nominal inputs and states $\bar{\mathbf u}^\star_{\mathbf N}, \bar{\mathbf x}^\star_{\mathbf N}$, via conditions \eqref{eq:zon_incl_compact_b}--\eqref{eq:zon_incl_compact_e}, \eqref{eq:zon_incl_compact_g}--\eqref{eq:zon_incl_compact_h}, ensure that \eqref{eq:one-step}--\eqref{eq:input-adm} hold. Furthermore, enforcing terminal constraint \eqref{eq:zon_incl_compact_i} and penalizing terminal cost $V_T(\bar x_N, \delta_N)$ in \eqref{eq:zon_incl_compact_a} guarantees \eqref{eq:terminal_invariance}--\eqref{eq:terminal_contractiveness}. \hfill\hfill\qed
\end{pf}

\begin{remark}
State and input in hyper-rectangles $\mathcal{X} = \{x\in \mathbb{R}^n: \underline{x} \leq x \leq \overline{x}\}$ admit both a zonotopic representation $\langle (\overline{x} +\underline{x})/2, \text{diag}(\overline{x} - \underline{x})/2 \rangle$ and a half-space representation $\{[I_n \, -I_n]^\top x \leq \mathbf{1}_{2n}\}$. The half-space form for \eqref{eq:zon_incl_compact_d}--\eqref{eq:zon_incl_compact_e} is adopted, as zonotopic encodings offer limited benefit for the low-order sets considered. For the terminal set \eqref{eq:zon_incl_compact_i}, a polyhedral representation remains necessary as it is generally non-zonotopic. Conversely, for \eqref{eq:zon_incl_compact_g}--\eqref{eq:zon_incl_compact_h}, polyhedral representations are strongly discouraged. \hfill \qed
\end{remark}

\begin{remark}
Unlike in the polyhedral case, treating the zonotope centers as independent variables requires explicitly enforcing the inclusion condition $c_k \in \langle 0, G \Delta_k\rangle$, which, in turn, ensures $\bar x_k \in \langle c_k, G\Delta_k\rangle$. \hfill \hfill \qed
\end{remark}
\begin{remark}
\label{rem:zon_homothetic}
The elastic tube MPC problem \eqref{eq:zon_incl_compact} reduces to the homothetic case by enforcing all components of the scaling vector to be equal, $\delta_k=\delta_k\cdot \mathbf 1_D$, $k \in \mathcal{I}_{[0,N]}$. \hfill \hfill \qed
\end{remark}

The central feature of Thm.~\ref{thm:elastic-tube-Phi} is the use of Cor.~\ref{cor:scaled_zon_condition} to enforce the one-step reachability condition \eqref{eq:one-step}. The setting is referred to as the zonotopic-elastic case. For comparison, two variants are introduced to replace \eqref{eq:zon_incl_compact_g}–\eqref{eq:zon_incl_compact_h}:
\begin{enumerate}[label=\roman*)]
    \item zonotopic-elastic with $\Phi_0$: the sequence $\{\Phi_k\}_{k \in \mathcal I_{[0, N-1]}}$ is replaced by a constant matrix $\Phi_0$, pre-computed according to \eqref{eq:Phi_0_computation};

    \item zonotopic-elastic with $\Gamma$: the matrix $\Gamma$ is pre-computed by solving a problem analogous to \eqref{eq:Phi_0_computation} tailored to set-inclusion conditions of the form \eqref{eq:zon_inclusion_test}.
\end{enumerate}

A preliminary version of the zonotopic-elastic scheme with $\Gamma$ was reported in \cite{DiaconescuEtAlCDC24}, although zonotope centers were not treated as decision variables and the reachability condition \eqref{eq:one-step} was imposed in its more restrictive equality form.

\begin{table}[!ht]
        \begin{center}
        \scalebox{0.815}{
        \begin{tabular}{|c|c|c|c|c|c|c|}
             \hline \multirow{2}{*}{Set}&\multirow{2}{*}{Tube}&\multicolumn{3}{|c|}{$\#$ Variables}&\multirow{2}{*}{$\#$ Inequalities}&\multirow{2}{*}{$\#$ Equalities}\\\cline{3-5}
             &&Decision&Scaling&Auxiliary&&\\\hline
             \multirow{3}{*}{$\mathcal{P}$}
             &$\mathcal R$&\multirow{12}{*}{$(N-1)m+Nn$}&0&\multirow{3}{*}{0}&\multirow{3}{*}{$(N-1)\left[2(n+m)+q\right]+q_T+\begin{cases}0\\N\\Nq\end{cases}$}&\multirow{3}{*}{0}\\\cline{2-2}\cline{4-4}
             &$\mathcal H$&&$N$&&&\\\cline{2-2}\cline{4-4}
             &$\mathcal E$&&$Nq$&&&\\\cline{1-2}\cline{4-7}
             \multirow{3}{*}{$\mathcal{Z}$ + c + $\Gamma$}
             &$\mathcal R$&&0&\multirow{3}{*}{$Nn + 2D(N-1)$}&\multirow{3}{*}{$(N-1)\left[2(n+m)+4D\right]+q_T+\begin{cases}0\\N\\ND\end{cases}$}&\multirow{3}{*}{$2n(N-1)$}\\\cline{2-2}\cline{4-4}
             &$\mathcal H$&&$N$&&&\\\cline{2-2}\cline{4-4}
             &$\mathcal E$&&$ND$&&&\\\cline{1-2}\cline{4-7}
             \multirow{3}{*}{{$\mathcal{Z}$ + c + $\Phi$}}
             &$\mathcal R$&&0&\multirow{3}{*}{$Nn + D (N-1) (2\bar{D} + 1)$}&\multirow{3}{*}{$(N-1)\left[2(n+m)+3D + 2D\bar{D}\right]+q_T+\begin{cases}0\\N\\ND\end{cases}$}&\multirow{3}{*}{$n(N-1)(\bar{D}+1)$}\\\cline{2-2}\cline{4-4}
             &$\mathcal H$&&$N$&&&\\\cline{2-2}\cline{4-4}
             &$\mathcal E$&&$ND$&&&\\\cline{1-2}\cline{4-7}
             \multirow{3}{*}{$\mathcal{Z}$ + c + $\Phi_{0}$}&$\mathcal R$&&0&\multirow{3}{*}{$Nn + 3D(N-1)$}&\multirow{3}{*}{$(N-1) [2(n+m) +5D]+q_T+\begin{cases}0\\N\\ND\end{cases}$}&\multirow{3}{*}{$2n(N-1)$}\\\cline{2-2}\cline{4-4}
             &$\mathcal H$&&$N$&&&\\\cline{2-2}\cline{4-4}
             &$\mathcal E$&&$ND$&&&\\\hline
        \end{tabular}}
        \end{center}
        \caption{Generic complexity 
        for various tube-MPC implementations}
        \label{table:complexity}
\end{table}

\section{Complexity analysis}
\label{sec:complexity}
Although the problem's size, measured in terms of the number of decision variables and constraints, does not provide a definitive predictor of the resulting computation time, it remains a valuable metric for a comparative analysis. Table~\ref{table:complexity} summarizes various scenarios for this purpose. Set types are indicated as $\mathcal{P}$ (polyhedron) or $\mathcal{Z}$ (zonotope), with centers treated as auxiliary optimization variables in all zonotopic cases. The presence of parameters $\Gamma$, $\Phi$, and $\Phi_0$ corresponds to the application of Lem.~\ref{lem:zon_inclusion}, Cor.~\ref{cor:scaled_zon_condition}, and Cor.~\ref{cor:Phi0}. Tube types are denoted by $\mathcal{R}$ (rigid), $\mathcal{H}$ (homothetic), and $\mathcal{E}$ (elastic). The analysis assumes hyper-rectangular constraints on states and inputs, represented by $2n$ and $2m$ inequalities, respectively. The variables $q$ and $q_T$ denote, respectively, the number of inequalities associated with the RPI set and the terminal set, as defined in Prop.~\ref{prop:rpi_approx_new} and \eqref{eq:mpi_rec_standard}. The shorthand notation $\bar{D} = D + D_{\omega} + 1$ is as first introduced in Prop.~\ref{prop:rpi_approx_new}. Parameter $N$ stands for the length of the MPC prediction horizon. In all zonotopic variants considered, the case `with centers' introduces an additional $(N-1)(n+\{0, 1, D\})$ variables, $n(N-1)$ equalities and $2 \cdot \{0, 1, D\}$ inequalities. The increased complexity is needed to maintain equivalence with the polyhedral cases, where the centers are optimized `by default'.

\emph{Tube profile complexity}. First, note that the RPI set may be represented either in its zonotopic form, with $D$ generators, or in its polyhedral form, to which, via Buck's formula \cite[Thm. 9.4]{fukuda2020polyhedral}, $q=2{D\choose n-1}\gg D$ half-spaces. This renders the polyhedral elastic representation from \cite{RakovicElastic2016} impractical for $n\geq 3$ and, also, for a larger order (ratio $D/n$), due to the increase in the number of variables and constraints. Furthermore, the zonotopic variations require substantially fewer scaling factors than those needed in the standard polyhedral elastic tube construction, as seen in Table~\ref{tab:scaling_factors}.
~\vspace{20em}\\

\begin{table}[!ht]
    \centering
    \begin{tabular}{l|c|c}
         Case & $\#$ Representation & $\#$ Scaling factors\\\hline
         $\mathcal P$ + $\mathcal E$& $q=2{D\choose n-1}$ & $q=2{D\choose n-1}$\\\hline
         $\mathcal Z$ + $\mathcal E$ (all mtds.)& $D$& $D$\\
    \end{tabular}
    \caption{No. of constraints and scaling factors}
    \label{tab:scaling_factors}
\end{table}


\emph{Inclusion test.} Beyond this, the differences are due to the number and nature of the auxiliary variables with `$\Gamma$' and `$\Phi_0$' formulations having similar complexity and both being significantly more compact than the `$\Phi$' case due to the presence of the product $D\bar D$. This drawback is, arguably, alleviated by the increased flexibility (recall that $\Gamma,\Phi_0$ are pre-computed, whereas $\Phi$ is part of the optimization problem). The number of constraints exhibits the same pattern, with the $\Phi$ case having significantly more of them.

\emph{Scaling method.} Both rigid and homothetic cases use similar scaling, regardless of the description used ($\mathcal{P}$ or $\mathcal{Z}$). Still, small differences may be observed in the number of auxiliary constraints and inequalities. As expected, the larger structural differences appear among the elastic methods, with several salient points: i) $\mathcal{P}$+$\mathcal{E}$ quickly fails due to $q\gg D$ and negatively compares even with $\mathcal{Z}$+$\mathcal{E}$$(\Phi)$, whose number of variables is $\approx ND^2$; ii) $\mathcal{Z}$+$\mathcal{E}$$(\Phi)$ avoids pre-computation of auxiliary variables ($\Gamma$ or $\Phi_0$) used in checking the set inclusions, at the price of increased complexity w.r.t. $\mathcal{Z}$+$\mathcal{E}$$(\Gamma)$ and $\mathcal{Z}$+$\mathcal{E}$$(\Phi_0)$. 

\emph{Terminal set design.} The design choices for the RPI and terminal sets significantly influence the overall complexity. The former is the result of an $s$-step iteration ($D \approx s D_{\omega}$, see Rem.~\ref{rem:complexity_vs_CDC}) and the latter depends on the specific case considered. Although, for simplicity, Table~\ref{table:complexity} reports a single parameter $q_T$, in reality $q_T$ depends jointly on the tube type $(\mathcal{R}/\mathcal{H}/\mathcal{Z})$ and the terminal set representation $(\mathcal{P}/\mathcal{Z})$. This dependence affects both the dimension of the terminal dynamics and the number of iterations required to verify the stopping condition for \eqref{eq:mpi_rec_standard}, which can lead to a rapid growth in the inequalities.

\section{Simulation and results} \label{sec:apps}
The implementation is illustrated for the double integrator from \cite{RakovicElastic2016} and the CSE example from \cite{Leibfritz2006COMPleibCM}. The optimizations are solved in MATLAB, using YALMIP \cite{Lofberg2012} and GUROBI, on a computer with a 2.1GHz i7 Intel processor with 12 cores and 32GB RAM, while the zonotopic sets are implemented with CORA \cite{Althoff2015ARCH}.

\subsection{Double integrator \cite{RakovicElastic2016}}
\label{apps:int}
\begin{figure}[!ht]
    \includegraphics[width=.8\columnwidth]{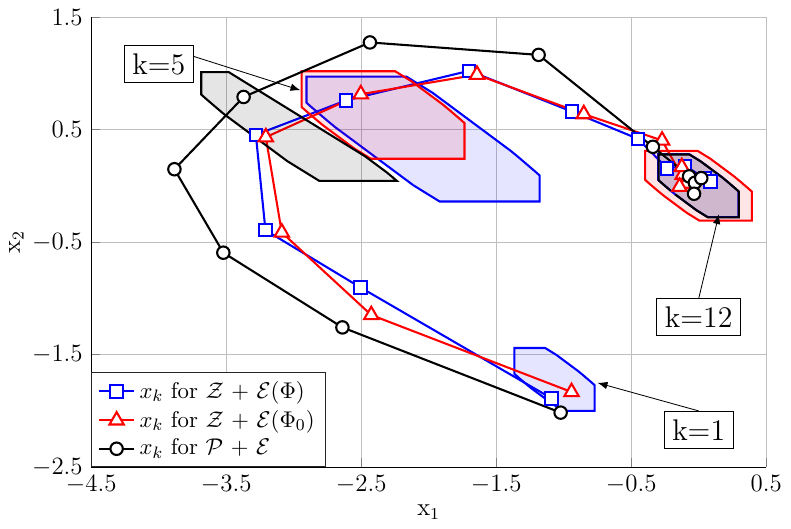}
    \caption{Elastic tube variants for the double integrator}
    \label{fig:tubes_double_integrator}
\end{figure}

The double integrator from \cite{RakovicElastic2016} is revisited, resulting in the zonotopic constructions from Fig.~\ref{fig:tubes_double_integrator} for three tube variants: polyhedral elastic as in \cite{RakovicElastic2016} (\begin{tikzpicture}[baseline=-2pt]\draw[black,line width=1pt,solid](0,0)--(.8,0); \node[draw,minimum width=1pt,line width=1pt,black,fill=white, circle, scale=0.75] at(0.4,0){};\end{tikzpicture}), $\mathcal{Z}$+$\mathcal{E}$($\Phi_0$) as in Cor.~\ref{cor:Phi0} (\begin{tikzpicture}[baseline=-2pt]\draw[red,line width=1pt,solid](0,0)--(.8,0); \node[draw,minimum width=1pt,line width=1pt,red,fill=white, regular polygon, regular polygon sides=3, scale=.5] at(0.4,0){};\end{tikzpicture}), and $\mathcal{Z}$+$\mathcal{E}$($\Phi$) as in Cor.~\ref{cor:scaled_zon_condition} (\begin{tikzpicture}[baseline=-2pt]\draw[blue,line width=1pt,solid](0,0)--(.8,0); \node[draw,minimum width=1pt,regular polygon, regular polygon sides=4, line width=1pt,blue,fill=white,scale=.6] at(0.4,0){};\end{tikzpicture}).

Within an MPC setup with $Q = I$, $R = 0.01$, $N = 12$, 
$\mathcal{X} = [-10, 10] \times [-10, 2]$, 
$\mathcal{U} = [-1, 1]$, 
$\mathcal{T} = \{0\}$, and 
$\mathcal{W} = [-0.1, 0.1] \times [-0.1, 0.1]$, 
all tube variants yield feasible solutions (solid lines in the figure). 
For all cases, the inclusion of a disturbance-affected trajectory within its corresponding tube is verified; instances $k \in \{1,5,12\}$ of the tube profile 
are highlighted to illustrate the stretching induced by the associated $\delta_k$ scaling factors.

\begin{figure}[!ht]
    \centering
    \includegraphics[width=0.8\columnwidth]{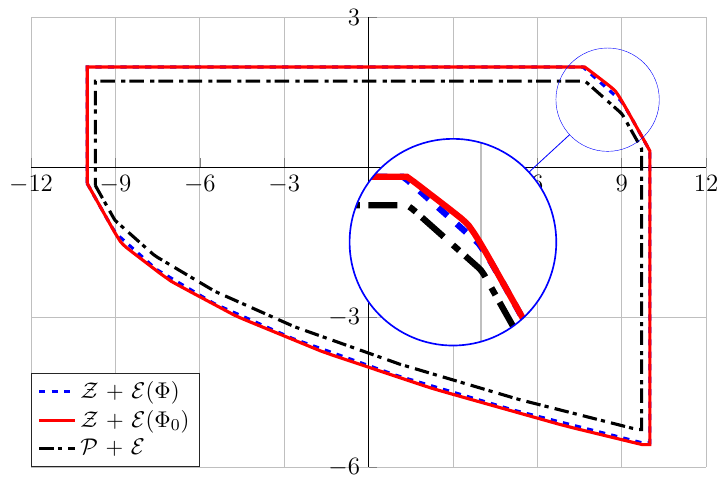}    
    \caption{Domain of attraction}
    \label{fig:doa_elastic}
\end{figure}

Evaluating the method's performance based on variations in the tube profile is often difficult to interpret, as it depends on the interplay between the choice of generator seeds, the penalization of homothetic or elastic scaling factors, and the imposed stage and terminal constraints. Therefore, it is not straightforward to categorically determine that a specific set size guarantees the best performance. A more unambiguous criterion is the size of the domain of attraction (DoA), defined as the set of initial states $\bar{x}_0$ for which feasible solutions exist. By definition, the DoA is obtained by projecting the feasible set of \eqref{eq:MPC_problem_nominal} onto the subspace of $\bar{x}_0$. Both this projection and the subsequent volume computation have exponential complexity and become intractable surprisingly quickly~\cite{chalkis2025volesti}. To address this, an iterative gridding procedure is employed to obtain inner and outer approximations of the DoA, with the process terminating once the difference falls below $1\%$. The resulting approximations for the same cases as before are shown in Fig~\ref{fig:doa_elastic}.

The results show that the zonotopic variants outperform the polyhedral case, with $\mathcal{Z}$+$\mathcal{E}$($\Phi$) exhibiting a slight advantage over the more restrictive $\mathcal{Z}$+$\mathcal{E}$($\Phi_0$), at the price of a higher computational burden. This difference is expected to become more pronounced in higher dimensions; in the $\mathbb{R}^2$ case shown here, the limited degrees of freedom constrain the potential improvement. 

\begin{figure*}[!ht]
    \centering
    \begin{minipage}[b]{0.48\textwidth}
        \centering
        \includegraphics[width=.9\textwidth]{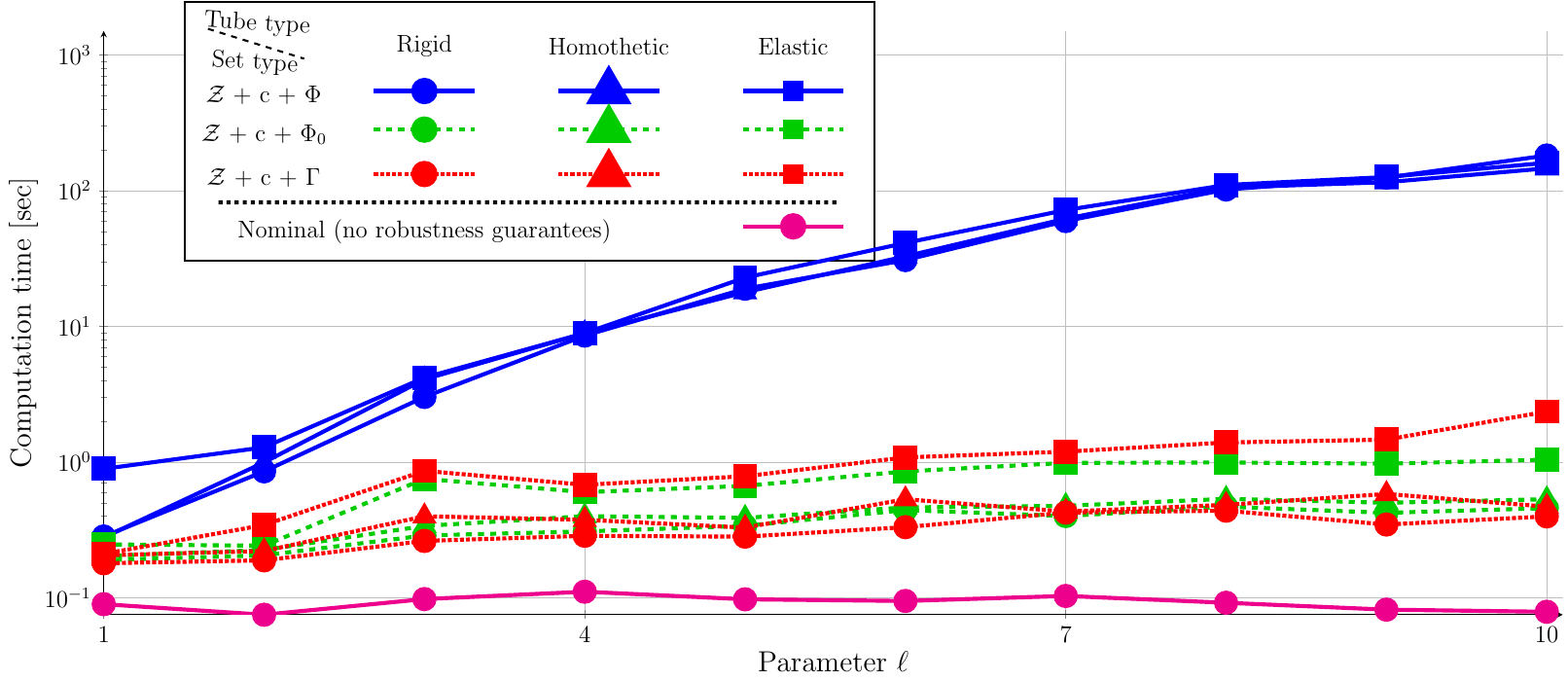}
        \caption{Solver runtimes for tube and set types with centers}
        \label{fig:zon_computation_times_centers}
    \end{minipage}
    \hfill
    \begin{minipage}[b]{0.48\textwidth}
        \centering
        \includegraphics[width=.9\textwidth]{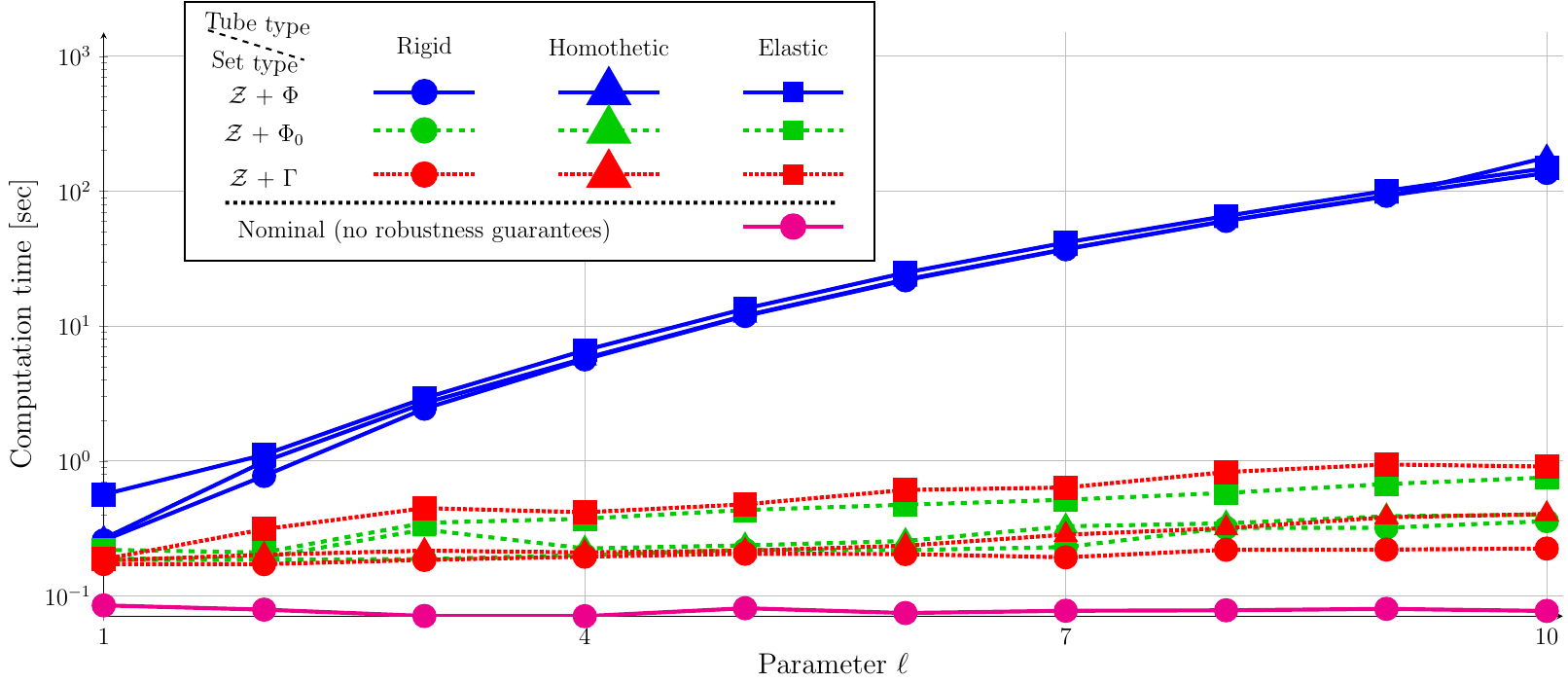}
        \caption{Solver runtimes for tube and set types without centers}
        \label{fig:zon_computation_times_nocenters}
    \end{minipage}
\end{figure*}
\subsection{Coupled Spring Experiment - CSE \cite{Leibfritz2006COMPleibCM}}
\label{apps:CSE}
The double integrator example is helpful for visualization purposes, but it cannot provide sufficient insight for computational effort quantification. The influence of the problem's size is assessed using the continuous-time CSE dynamics, modeled as a system of interconnected springs, dampers, and masses, as provided in $COMPl_{e}ib$ \cite{Leibfritz2006COMPleibCM}. The system states are defined by the positions and velocities of the masses, while the inputs correspond to the two forces applied at the ends of the spring chain. The parameter $\ell$ represents the number of springs and determines the model size ($2\ell$), as given by 
\vspace{-1em}
\begin{align}
\label{eq:cse1}
    \dot x(t) &= 
    \underbrace{\begin{bmatrix}
    0 & I \\
    -M_{c}^{-1}K_{c} & -M_{c}^{-1}L_{c}
    \end{bmatrix}}_{\text{A}\in \mathbb R^{2\ell \times 2\ell}}x(t) + 
    \underbrace{\begin{bmatrix}
    0 \\
    M_{c}^{-1}D_{c}
    \end{bmatrix}}_{\text{B}\in \mathbb R^{2\ell \times 2}} u(t) \,,
\end{align}
where\vspace{-3em}
\begin{align}
    K_{c} = k\begin{bmatrix}
    1 & -1 & \cdots & 0 & 0 \\
    -1 & -2 & \ddots & 0 & 0 \\
    \vdots & \ddots & \ddots & \ddots & \vdots \\
    0 & 0 & \ddots & -2 & -1 \\
    0 & 0 & \cdots & -1 & 1
    \end{bmatrix}, \nonumber \; D_{c} = \begin{bmatrix} \;
    1 & \hphantom{-}0 \\
    0 & \hphantom{-}0 \\
    \vdots & \hphantom{-}\vdots \\
    0 & \hphantom{-}0 \\
    0 & -1
    \end{bmatrix}, \nonumber
\end{align}
and $M_{c} = \mu I$, $L_{c} = \tau I$.

The parameters are $\mu = 4, \; \tau = 1, \; k = 1$, and the continuous system is discretized with the forward Euler method for a sampling of $1$ sec. State and input constraints are $\mathcal X= [-1, 1]^{2\ell},\:  \mathcal U=[-1, 1]^{2},\: \mathcal W=[-0.001, 0.001]^{2\ell}$, while the MPC parameters are $Q=I_{2\ell}$, $R=0.01I_2$, $N=25$.

All tube polyhedral set implementations fail for approximately $\ell \geq 3$, either due to exceeding the computational time limit, reaching the maximum allowed number of iterations or by not finding a feasible solution, whereas the zonotopic implementations remain successful across the entire range of parameters, up to $20$ states. 

Figs.~\ref{fig:zon_computation_times_centers}and~\ref{fig:zon_computation_times_nocenters} report the computation times for 18 zonotopic variants -- combinations of method type ($\mathcal{R}$/$\mathcal{H}$/$\mathcal{E}$), auxiliary variables ($\Gamma$/$\Phi$/$\Phi_0$), and with/without centers -- together with the nominal case. The latter cannot guarantee that the predicted trajectory remains within the prescribed tube profile for all disturbance realizations $\omega_k \in \mathcal{W}$ over the prediction horizon, and it is included here solely to indicate the additional computational overhead of the tube-based methods.

The set-containment conditions formulated with zonotopic sets exhibit linear, and quadratic in the case of $\Phi$, computational complexity. Furthermore, the requirement to preselect a value for $\Gamma$ \cite{DiaconescuEtAlCDC24}, originally imposed to circumvent a nonlinear optimization problem, is mitigated through the adoption of the $\Phi$-variant, for which the computational trade-off are explicitly quantified.

The polyhedral method inherently optimizes the tube centers, whereas, in the zonotopic framework, this aspect is primarily a design choice. Computation times corresponding to all considered variants are presented in Figs.~\ref{fig:zon_computation_times_centers} and \ref{fig:zon_computation_times_nocenters}. The augmentation of the optimization problem with center variables results in a marginal computational overhead, reflecting the expanded dimensionality of the decision space.

Additionally, no acceleration techniques were employed. Further investigations should include vectorization and parallelization techniques along with the usage of high-performance solvers like Tenscalc \cite{Tenscalc2022} or OSQP \cite{osqp}. Two mitigating factors are noteworthy: despite the increased computation time from additional variables, tube variants provide crucial robustness guarantees; furthermore, this overhead is less critical if the MPC problem is solved offline for trajectory generation.

\vspace{-3mm}





\section{Conclusions} \label{sec:conclusions}
The integration of elastically-scaled zonotopes into elastic-tube MPC significantly reduces runtime and improves scalability for high-dimensional systems. Validated on benchmark models, the approach effectively manages additive disturbances in complex scenarios. Future work will address the sensitivity of tube-based methods to the initial robust positively invariant set by reducing its complexity. Additionally, this model-based framework can be extended to data-driven MPC \cite{Alanwar2024}.


\section*{Acknowledgments}

The authors would like to thank Andrei Sperilă and Zhizhong Guo for insightful discussions on the subject.

\end{document}